\documentclass[preprintnumbers,superscriptaddress,pra,showkeys]{revtex4}
\usepackage{amsfonts}
\usepackage{amssymb}
\usepackage{amsmath}
\usepackage{epsfig}
\usepackage{graphicx}
\usepackage{color}

\begin{document}

\title{The curvature-induced gauge potential and the geometric momentum for
a particle on a hypersphere }
\author{Z. Li}
\affiliation{School for Theoretical Physics, School of Physics and Electronics, Hunan
University, Changsha 410082, China}
\author{L. Q. Lai}
\affiliation{School for Theoretical Physics, School of Physics and Electronics, Hunan
University, Changsha 410082, China}
\author{Y. Zhong}
\email{zhongy@hnu.edu.cn}
\affiliation{School for Theoretical Physics, School of Physics and Electronics, Hunan
University, Changsha 410082, China}
\author{Q. H. Liu}
\email{quanhuiliu@gmail.com}
\affiliation{School for Theoretical Physics, School of Physics and Electronics, Hunan
University, Changsha 410082, China}
\date{\today }

\begin{abstract}
A particle that is constrained to freely move on a hyperspherical surface in
an $N\left( \geq 2\right) $ dimensional flat space experiences a
curvature-induced gauge potential, whose form was given long ago (J. Math.
Phys. \textbf{34}(1993)2827). We demonstrate that the momentum for the
particle on the hypersphere is the geometric one including the gauge
potential and its components obey the commutation relations $\left[
p_{i},p_{j}\right] =-i\hbar J_{ij}/r^{2}$, in which $\hbar $ is the Planck's
constant, and $p_{i}$ ($i,j=1,2,3,...N$) denotes the $i-$th component of the
geometric momentum, and $J_{ij}$ specifies the $ij-$th component of the
generalized\textit{\ angular momentum} containing both the orbital part and
the coupling of the generators of continuous rotational symmetry group $%
SO(N-1)$ and curvature, and $r$ denotes the radius of the $N-1$ dimensional
hypersphere.
\end{abstract}

\maketitle

\section{Introduction}

In quantum mechanics, a constrained dynamical system is usually associated
with a gauge structure, and it is quite well-understood in, for instance,
gravitational field \cite{RMP1957,Hehl}, condensed matter physics \cite%
{2010PR,Iorio}, quantum fields \cite{weinberg} and particle physics \cite%
{Qiugg}. We are recently interested in the constrained motion, i.e., a
particle remains and freely moves on a hypersurface \cite%
{liu07,liu11,liu13,liu132,liu133,liu14,liu15,liu16,liu18,liu19,liu20,liu21}.
There are also a lot of papers paying attention to the curvature-induced
gauge structure for the system, and the form of the gauge potential is
well-known \cite%
{japan1993,japan1993-2,japan1995,italy1995,japan1997,japan1997-2,japan1997-3,japan1998,usa2003,pra14,soiqm,japan2007,yang}%
. The form of the gauge potential is clearly a coupling of the curvature and
all generators of a rotational symmetry group, which in quantum mechanics is
not necessarily understood to represent the spin though it is the important
situation. Even so, the gauge potential is far beyond fully understood. For
instance, once a classical bracket (c.f. Eq. (\ref{ppL})) involves both the
momentum and the orbital angular momentum, the corresponding quantum
commutator (c.f. Eq. (\ref{ppJ})) based on the quantization of the classical
bracket contains both the momentum operator and the angular momentum
operator, and it appears transparent. However, the quantum commutator is
highly non-trivial because the momentum explicitly contains the extrinsic
curvature, and the previous studies unintentionally missed this geometric
nature of the momentum till 2011 when the geometric momentum came into sight 
\cite{liu11}. The main aim of the present study is to show that the relation
(\ref{ppJ}) holds true with reasonable inclusion of the gauge potential into
the geometric momentum.

This paper is organized in the following. In section II, how Ohnuki and
Kitakado obtained the gauge potential is outlined and commented. In section
III, the Dirac formalism of quantization for the particle on the sphere is
invoked, where the Dirac brackets between the momentum components and the
orbital angular momentum components play central role. To show that the
fundamental quantum conditions in the Dirac scheme of quantization are
completely compatible with the gauge potential, we must utilize the proper
form of the momentum. In section IV and V, we explicitly prove two sets of
fundamental quantum conditions in the Dirac scheme, respectively. The final
section VI is a brief conclusion and discussion.

\section{Ohnuki and Kitakado gauge potential on hyperspherical surface}

By a hyperspherical surface $\mathit{S}^{N-1}$ ($N\geqslant 2$), we mean a
spherical surface in $N$ dimensional flat space $E^{N}$. The surface
equation is given by, 
\begin{equation}
x_{i}x_{i}=r^{2},(i=1,2,3,...,N),\text{ and }n_{i}=\frac{x_{i}}{r},
\label{S}
\end{equation}%
where $r$ denotes the radius of the sphere, and $x_{i}$ is the $i$-th
coordinate and $n_{i}$\ is the $i$-th component of the normal vector.
Throughout the paper, the Einstein's summation convention is adopted, which
implies the indices repeated twice in a term are summed over the range of
the index unless specified, and the Roman indices $i,j,k,l,m,n$ run from $1$
to $N$ to specify the coordinates of the surface in $E^{N}$, while the Greek
ones $\alpha ,\beta ,\lambda ,\mu $ run from $1$ to $N-1$ which can not be
confused with the $N-1$ local coordinates of the surface.

Ohnuki and Kitakado\ started from a \emph{fundamental algebra} on the flat $%
E^{N}$ space \cite{japan1993}, the corresponding Euclidean group $E(N)$
whose generators $\left\{ x_{i},J_{jk}\right\} $ satisfy following
relations, 
\begin{eqnarray}
\lbrack x_{i},x_{j}] &=&0,\left[ x_{k},J_{ij}\right] =i\hbar \left(
x_{i}\delta _{kj}-x_{j}\delta _{ki}\right) ,  \label{OK1} \\
\left[ J_{jk},J_{lm}\right]  &=&i\hbar \left( \delta _{jl}J_{km}-\delta
_{jm}J_{kl}+\delta _{km}J_{jl}-\delta _{kl}J_{jm}\right) .  \label{OK2}
\end{eqnarray}%
The $SO(N)$ (\ref{OK2}) is a subgroup of $E(N)$, whose generators are the 
\textit{generalized angular momentum}, 
\begin{equation}
J_{ij}=L_{ij}+f_{ij},  \label{J}
\end{equation}%
where $L_{ij}$ $=x_{i}(-i\hbar \partial _{j})-x_{j}(-i\hbar \partial _{i})$
is $ij-$th component of the usual orbital angular momentum, and $f_{ij}$ is
related to $S_{\alpha \beta }$ that obey commutation relations (\ref{OK2})
as well and constitute the irreducible representation of $SO(N-1)$ group 
\cite{japan1993}, 
\begin{eqnarray}
f_{\alpha \beta } &=&-f_{\beta \alpha }=S_{\alpha \beta },  \label{f1} \\
f_{\alpha N} &=&-f_{N\alpha }=-\frac{r}{r+x_{N}}S_{\alpha \beta }n_{\beta }.
\label{f2}
\end{eqnarray}%
Ohnuki and Kitakado obtained the form of $f_{ij}$ by means of consideration
of the transportation of a state from any point to another on $\mathit{S}%
^{N-1}$ by successive infinitesimal unitary transformation \cite{japan1993}.
The gauge structure is rooted in a coupling between the generators $%
S_{\alpha \beta }$ of rotational symmetry group $SO(N-1)$ and curvature via
relation (\ref{f2}). To be\ explicit, the general form of the gauge
potential $\mathbf{A}$ is in the following \cite{japan1993},%
\begin{equation}
A_{i}=\frac{1}{r}f_{ij}n_{j}=\frac{1}{r+x_{N}}\left\{ 
\begin{array}{cc}
S_{i\beta }n_{\beta }, & i\neq N \\ 
0, & i=N%
\end{array}%
\right. .  \label{A}
\end{equation}

In the following, we point out three elementary properties of the gauge
field $\mathbf{A}$. 1. Commutators $\left[ A_{\alpha },A_{\beta }\right] $
between the two different components of the gauge potential $\mathbf{A}$
indicate that the gauge field is non-Abelian, 
\begin{subequations}
\begin{eqnarray}
\left[ A_{\alpha },A_{\beta }\right]  &=&\frac{n_{\alpha }n_{\beta }}{\left(
r+x_{N}\right) ^{2}}\left[ S_{\alpha \mu },S_{\beta \nu }\right] 
\label{AA1} \\
&=&\frac{i\hbar }{\left( r+x_{N}\right) ^{2}}\left( \left( 1-\frac{x_{N}^{2}%
}{r^{2}}\right) S_{\alpha \beta }+\left( n_{\alpha }S_{\beta \mu }-n_{\beta
}S_{\alpha \mu }\right) n_{\mu }\right)   \label{AA2} \\
&=&i\hbar \left( \frac{\left( r-x_{N}\right) }{r^{2}\left( r+x_{N}\right) }%
S_{\alpha \beta }+\frac{1}{r+x_{N}}\left( n_{\alpha }A_{\beta }-n_{\beta
}A_{\alpha }\right) \right)   \label{AA3} \\
&=&i\hbar \left( \frac{\left( r-x_{N}\right) }{r^{2}\left( r+x_{N}\right) }%
S_{\alpha \beta }+\frac{1}{r+x_{N}}\left( \mathbf{n}\wedge \mathbf{A}\right)
_{\alpha \beta }\right) \neq 0.  \label{AA4}
\end{eqnarray}%
The field strength and the gauge transformation of the gauge field $\mathbf{A%
}$ are discussed in Ref. \cite{japan1997-3}.

2. It is easily to check that the $N$-th component of the gauge potential $%
\mathbf{A}$ is zero, i.e., $A_{N}=n_{\alpha }S_{\alpha \beta }n_{\beta
}/(r+x_{N})=0$, which amounts to choose a convenient gauge condition. More
importantly, the gauge potential $\mathbf{A}$ is defined on the tangential
plane on $\mathit{S}^{N-1}$ for we have, 
\end{subequations}
\begin{equation}
\mathbf{A\cdot n}=0\text{, or, }A_{\alpha }n_{\alpha }=0.  \label{An}
\end{equation}

3. A very important case is: Once $S_{\alpha \beta }$ is understood as spin
matrix, the gauge field $\mathbf{A}$ can be interpreted as spin-curvature
coupling, where the curvature part $n_{\beta }/\left( r+x_{N}\right) $ in (%
\ref{A}) is related to spin-connections, which is in detail discussed in
Ref. \cite{japan1997-2}. Once the radius $r$ is infinitely large, we have $%
n_{\beta }/\left( r+x_{N}\right) \rightarrow 0$. Thus, the spin is not
directly the usual intrinsic angular momentum of the particle defined in
flat space.

\section{Dirac formalism of quantization on the sphere $\mathit{S}^{N-1}$
and an $SO(N,1)$ algebra}

Let us now consider following equation of the hyperspherical surface
equation,%
\begin{equation}
f\left( x\right) =\frac{1}{2r}\left( x_{i}^{2}-r^{2}\right) =0.  \label{s}
\end{equation}%
From it we know $\mathbf{n}=\nabla f$. Dirac formalism for constrained
motion gives following Dirac brackets \cite{liu13}, 
\begin{eqnarray}
\left[ x_{i},x_{j}\right] _{D} &=&0,  \label{xx} \\
\left[ x_{i},p_{j}\right] _{D} &=&\left( \delta _{ij}-n_{i}n_{j}\right) ,
\label{xp} \\
\left[ p_{i},p_{j}\right] _{D} &=&-\frac{L_{ij}}{r^{2}}=-\frac{%
x_{i}p_{j}-x_{j}p_{i}}{r^{2}},  \label{ppL}
\end{eqnarray}%
where $L_{ij}=x_{i}p_{j}-x_{j}p_{i}$ is the usual\ $ij$-component of\ the
orbital angular momentum satisfying, 
\begin{eqnarray}
\left[ x_{k},L_{ij}\right] _{D} &=&\left( x_{i}\delta _{kj}-x_{j}\delta
_{ki}\right) ,  \label{xl} \\
\left[ p_{k},L_{ij}\right] _{D} &=&\left( p_{i}\delta _{kj}-p_{j}\delta
_{ki}\right) ,  \label{pLp} \\
\left[ L_{jk},L_{lm}\right] _{D} &=&\left( \delta _{jl}L_{km}-\delta
_{jm}L_{kl}+\delta _{km}L_{jl}-\delta _{kl}L_{jm}\right) .  \label{LLL}
\end{eqnarray}

When transition to quantum mechanics, we utilize the full Dirac formalism of
quantization procedure to define a quantum commutator for any pair of\textrm{%
\ }operators $f$ and $g$ is given by $\left[ f,g\right] =i\hbar $ $O\left\{ %
\left[ f,g\right] _{D}\right\} $. Thus, the proper form of quantum
conditions from (\ref{xx}) to (\ref{LLL}) must be, together with (\ref{OK1})
and (\ref{OK2}), 
\begin{eqnarray}
\left[ x_{i},x_{j}\right]  &=&0, \\
\left[ x_{i},p_{j}\right]  &=&i\hbar \left( \delta _{ij}-n_{i}n_{j}\right) ,
\\
\left[ p_{i},p_{j}\right]  &=&-i\hbar \frac{J_{ij}}{r^{2}}\neq -\frac{i\hbar 
}{r^{2}}\left( x_{i}p_{j}-x_{j}p_{i}\right) ,  \label{ppJ} \\
\left[ p_{k},J_{ij}\right]  &=&i\hbar \left( p_{i}\delta _{kj}-p_{j}\delta
_{ki}\right) .  \label{pJp}
\end{eqnarray}%
Here, we replace the symbol $L_{ij}$ by $J_{ij}$, which denotes the
generalized angular momentum, then the quantum momentum $p_{i}$ in our
problem is not the usual one. It is clear that operators ($p_{i},J_{ij}$)
form the $SO(N,1)$ algebra from relations (\ref{OK2}), (\ref{ppJ}) and (\ref%
{pJp}). 

We will show that with $\mathbf{p}=\mathbf{\Pi -A}$, relations (\ref{ppJ})
and (\ref{pJp}) are valid, in which the $\mathbf{A}$-independent part $%
\mathbf{\Pi }$ is a well-defined geometric momentum \cite%
{liu07,liu11,liu13,liu132,liu133,liu14,liu15,liu16,liu18,liu19,liu20,liu21}.
In general, the geometric momentum takes the form, $\mathbf{\Pi =}\Pi _{i}%
\mathbf{e}_{i}=$ $-i\hbar ({\nabla _{\Sigma }}+{M{\mathbf{n/}}}2)$ with $%
\nabla _{S}\equiv \nabla -\mathbf{n}\left( \mathbf{n\cdot }\nabla \right)
=\nabla -\mathbf{n}\partial _{n}$ denoting the gradient operator on $\mathit{%
S}^{N-1}$, and the mean curvature ${M\equiv -\nabla _{\Sigma }}\cdot \mathbf{%
n}$ is defined by the sum of the all principal curvatures. Notice that the
mean curvature ${M}$ is an extrinsic curvature, this form of momentum $%
\mathbf{p}=\mathbf{\Pi -A}$ is fundamentally different from the canonical
ones in curvilinear coordinates for $\mathbf{p}$ depends on the geometric
invariants whereas the the canonical ones do not. However, for a particle on
the hyperspherical surface, the geometric momentum assumes a simpler form
which can be obtained via a simple quantization of relation $%
p_{i}=n_{j}L_{ji}/r$ which in quantum mechanics turns out to be, 
\begin{equation}
\Pi _{i}=\frac{1}{2r}\left( n_{j}L_{ji}+L_{ji}n_{j}\right) ,\text{ i.e., }%
\mathbf{\Pi }=\mathbf{-}i\hbar \left( \nabla _{S}-(N-1)\frac{\mathbf{n}}{2r}%
\right) ,  \label{pJ}
\end{equation}%
where the fact $\left[ n_{j},L_{ji}\right] \neq 0$ is noted. Now we have $%
J_{ij}=L_{ij}+f_{ij}$ (\ref{J}). The similar quantization of the classical
relation $p_{i}=n_{j}L_{ji}/r$ gives explicitly how the gauge potential
enters in the geometric momentum, 
\begin{equation}
p_{i}=\frac{1}{2r}\left( n_{j}J_{ji}+J_{ji}n_{j}\right) =\Pi _{i}-A_{i},
\label{np}
\end{equation}%
where $A_{i}=f_{ij}n_{j}/r$ comes from its definition (\ref{A}). What is
more, the momentum still lies on the tangential plane on the surface for we
have,%
\begin{equation}
\mathbf{p\cdot n+n\cdot p}{\mathbf{=0,}}
\end{equation}%
where $A_{i}n_{i}=0$ directly arises from its definition (\ref{A}) as well,
and $\mathbf{\Pi \cdot n+n\cdot \Pi =0}$ is easily verifiable with
definition of $\mathbf{\Pi }$ (\ref{pJ}).

There is a curious fact. The form $\left[ p_{i},p_{j}\right]
_{D}=-L_{ij}/r^{2}$ (\ref{ppL}) is seldom used, and it is usually written in
the form $\left[ p_{i},p_{j}\right] _{D}=-\left(
x_{i}p_{j}-x_{j}p_{i}\right) /r^{2}$ whose quantization assumes the form $%
\left[ p_{i},p_{j}\right] =-i\hbar \left( x_{i}p_{j}-x_{j}p_{i}\right) /r^{2}
$, see for instance Weinberg's lectures on quantum mechanics \cite{Weinberg1}
and also Ohnuki and Kitakado paper \cite{japan1997-2}, in which only the
orbital angular momentum presents. However, if starting from $\left[
p_{i},p_{j}\right] _{D}=-L_{ij}/r^{2}$ (\ref{ppL}) which is explicitly
expressed in terms of orbital angular momentum, the quantum angular momentum
allows the spin-related angular momentum to enter because the angular
momentum in quantum mechanics is determined by the algebra. In other words,
the quantum conditions assume\ the form $\left[ p_{i},p_{j}\right] =-i\hbar
J_{ij}/r^{2}$ (\ref{ppJ}) in which $J_{ij}$ is known, then the proper form
the $p_{i}$ is sought. 

In section IV, we will show\ $\left[ p_{i},p_{j}\right] =-i\hbar J_{ij}/r^{2}
$ (\ref{ppJ}) with $p_{i}$ defined in (\ref{np}); and in section V, we will
prove $\left[ p_{k},J_{ij}\right] =i\hbar \left( p_{i}\delta
_{kj}-p_{j}\delta _{ki}\right) $ (\ref{pJp}). 

\section{A proof of $\left[ p_{i},p_{j}\right] =-i\hbar J_{ij}/r^{2}$}

We start from $\mathbf{p}=\mathbf{\Pi -A}$, from which the commutators $%
\left[ p_{i},p_{l}\right] $ are,%
\begin{equation}
\left[ p_{i},p_{l}\right] =\left[ \Pi _{i}-A_{i},\Pi _{l}-A_{l}\right] .
\label{start}
\end{equation}%
We split $N$ components $p_{i}$ of the momentum $\mathbf{p}$ into two
categories of $p_{\alpha }$ $\left( \alpha =1,2,\cdots ,N-1\right) $ and $%
p_{N}$, 
\begin{eqnarray}
p_{\alpha } &=&\Pi _{\alpha }+\frac{1}{r^{2}}x_{i}f_{i\alpha }=\Pi _{\alpha
}-A_{\alpha },  \label{pa} \\
p_{N} &=&\Pi _{N}-A_{N}=\Pi _{N},  \label{pn}
\end{eqnarray}%
where we used $A_{N}=0$. The calculations of $\left[ p_{i},p_{l}\right] $
will be done separately, and we first study $\left[ p_{\alpha },p_{\beta }%
\right] $ and secondly deal with $\left[ p_{\alpha },p_{N}\right] $.

In order to compute $\left[ p_{\alpha },p_{\beta }\right] $, we split the
commutator $\left[ p_{\alpha },p_{\beta }\right] $ into four parts, 
\begin{equation}
\left[ p_{\alpha },p_{\beta }\right] =\left[ \Pi _{\alpha }-A_{\alpha },\Pi
_{\beta }-A_{\beta }\right] =\left[ \Pi _{\alpha },\Pi _{\beta }\right] -%
\left[ \Pi _{\alpha },A_{\beta }\right] -\left[ A_{\alpha },\Pi _{\beta }%
\right] +\left[ A_{\alpha },A_{\beta }\right] .  \label{pp}
\end{equation}%
In the right-hand-side of above expression, the first part is known \cite%
{liu13} and the result is given by, 
\begin{equation}
\left[ \Pi _{i},\Pi _{j}\right] =-\frac{i\hbar }{r^{2}}L_{ij}.  \label{gmL}
\end{equation}%
The second\ and the third part is essentially the same for $\left[ \Pi
_{\alpha },A_{\beta }\right] =-\left[ A_{\beta },\Pi _{\alpha }\right] $,
and the last one is $\left[ A_{\alpha },A_{\beta }\right] $ which is also
understood via Eq. (\ref{AA3}). To carry out $\left[ \Pi _{\alpha },A_{\beta
}\right] $, we start from $\left[ \mathbf{\Pi },A_{\beta }\right] $ and then
project onto the $\alpha $-direction. To note a relation $\left[ \mathbf{\Pi 
},A_{\beta }\right] =-i\hbar \left[ \nabla -\mathbf{n}\left( \mathbf{n\cdot }%
\nabla \right) ,A_{\beta }\right] $, we need thus to compute the commutator $%
\left[ \nabla ,A_{\beta }\right] $, 
\begin{eqnarray}
\left[ \nabla ,A_{\beta }\right]  &=&\nabla A_{\beta }  \notag \\
&=&\nabla \left( \frac{n_{\mu }}{r+x_{N}}\right) S_{\beta \mu }  \notag \\
&=&\left( -\left( \frac{n_{\mu }\left( n_{N}+2\right) }{\left(
r+x_{N}\right) ^{2}}\right) \mathbf{n-}\left( \frac{n_{\mu }}{\left(
r+x_{N}\right) ^{2}}\right) \mathbf{e}_{N}+\left( \frac{1}{r\left(
r+x_{N}\right) }\right) \mathbf{e}_{\mu }\right) S_{\beta \mu }.  \label{DA}
\end{eqnarray}%
Then $\mathbf{n\cdot }\nabla A_{\beta }$ is with use of $\mathbf{n\cdot e}%
_{N}=n_{N}$ and $\mathbf{n\cdot e}_{\mu }=n_{\mu }$, 
\begin{equation}
\mathbf{n\cdot }\nabla A_{\beta }=\left( -\frac{2n_{\mu }\left(
n_{N}+1\right) }{\left( r+x_{N}\right) ^{2}}\mathbf{+}\frac{n_{\mu }}{%
r\left( r+x_{N}\right) }\right) S_{\beta \mu }.  \label{NA}
\end{equation}%
Thus, we obtain,%
\begin{eqnarray}
\left[ \nabla -\mathbf{n}\left( \mathbf{n\cdot }\nabla \right) ,A_{\beta }%
\right]  &=&\nabla A_{\beta }-\mathbf{n}\left( \mathbf{n\cdot }\nabla
\right) A_{\beta }  \notag \\
&=&\left( \left( \frac{n_{\mu }n_{N}}{\left( r+x_{N}\right) ^{2}}\mathbf{-}%
\frac{n_{\mu }}{r\left( r+x_{N}\right) }\right) \mathbf{n}-\left( \frac{%
n_{\mu }}{\left( r+x_{N}\right) ^{2}}\right) \mathbf{e}_{N}+\left( \frac{1}{%
r\left( r+x_{N}\right) }\right) \mathbf{e}_{\mu }\right) S_{\beta \mu } 
\notag \\
&=&\left( \mathbf{-}\frac{n_{\mu }}{\left( r+x_{N}\right) ^{2}}\mathbf{n}-%
\frac{n_{\mu }}{\left( r+x_{N}\right) ^{2}}\mathbf{e}_{N}+\frac{1}{r\left(
r+x_{N}\right) }\mathbf{e}_{\mu }\right) S_{\beta \mu }.  \label{PA}
\end{eqnarray}%
Projecting it onto the $\alpha $-direction, we get $\left[ \Pi _{\alpha
},A_{\beta }\right] $, with $\mathbf{e}_{\alpha }\cdot \mathbf{n=}$ $%
n_{\alpha }$, $\mathbf{e}_{\alpha }\cdot \mathbf{e}_{N}=0$, and $\mathbf{e}%
_{\alpha }\cdot \mathbf{e}_{\mu }=\delta _{\alpha \mu }$, 
\begin{eqnarray}
\left[ \Pi _{\alpha },A_{\beta }\right]  &=&-i\hbar \mathbf{e}_{\alpha
}\cdot \left[ \nabla -\mathbf{n}\left( \mathbf{n\cdot }\nabla \right)
,A_{\beta }\right]   \notag \\
&=&-i\hbar \left( \mathbf{-}\frac{n_{\mu }}{\left( r+x_{N}\right) ^{2}}%
n_{\alpha }+\frac{1}{r\left( r+x_{N}\right) }\delta _{\alpha \mu }\right)
S_{\beta \mu }  \notag \\
&=&-i\hbar \left( -\frac{n_{\alpha }}{r+x_{N}}A_{\beta }+\frac{S_{\beta
\alpha }}{r\left( r+x_{N}\right) }\right) .  \label{gma}
\end{eqnarray}%
So, two terms in right-hand-side of (\ref{pp}) $\left[ \Pi _{\alpha
},A_{\beta }\right] +\left[ A_{\alpha },\Pi _{\beta }\right] $ is,%
\begin{eqnarray}
\left[ \Pi _{\alpha },A_{\beta }\right] +\left[ A_{\alpha },\Pi _{\beta }%
\right]  &=&\left[ \Pi _{\alpha },A_{\beta }\right] -\left[ \Pi _{\beta
},A_{\alpha }\right]   \notag \\
&=&-i\hbar \left( -\frac{n_{\alpha }}{r+x_{N}}A_{\beta }+\frac{S_{\beta
\alpha }}{r\left( r+x_{N}\right) }-\left( -\frac{n_{\beta }}{r+x_{N}}%
A_{\alpha }+\frac{S_{\alpha \beta }}{r\left( r+x_{N}\right) }\right) \right) 
\notag \\
&=&-i\hbar \left( -\frac{1}{r+x_{N}}\left( n_{\alpha }A_{\beta }-n_{\beta
}A_{\alpha }\right) +\frac{2S_{\beta \alpha }}{r\left( r+x_{N}\right) }%
\right) .  \label{AA}
\end{eqnarray}%
Combining results (\ref{AA3}) and (\ref{AA}), we find a simple result for
last three commutators in (\ref{pp}), 
\begin{eqnarray}
-\left[ \Pi _{\alpha },A_{\beta }\right] -\left[ A_{\alpha },\Pi _{\beta }%
\right] +\left[ A_{\alpha },A_{\beta }\right]  &=&i\hbar \left( \frac{%
2S_{\beta \alpha }}{r\left( r+x_{N}\right) }-\frac{1}{r+x_{N}}\left(
n_{\alpha }A_{\beta }-n_{\beta }A_{\alpha }\right) \right)   \notag \\
&&+i\hbar \left( \frac{\left( r-x_{N}\right) }{r^{2}\left( r+x_{N}\right) }%
S_{\alpha \beta }+\frac{1}{r+x_{N}}\left( n_{\alpha }A_{\beta }-n_{\beta
}A_{\alpha }\right) \right)   \notag \\
&=&i\hbar \left( \frac{\left( r-x_{N}\right) }{r^{2}\left( r+x_{N}\right) }-%
\frac{2}{r\left( r+x_{N}\right) }\right) S_{\alpha \beta }  \notag \\
&=&i\hbar \left( -\frac{1}{r^{2}}+\left( \frac{1}{r^{2}}-\frac{1}{r\left(
r+x_{N}\right) }-\frac{x_{N}}{r^{2}\left( r+x_{N}\right) }\right) \right)
S_{\alpha \beta }  \notag \\
&=&-i\hbar \frac{S_{\alpha \beta }}{r^{2}}.  \label{last3}
\end{eqnarray}%
Thus we have from Eqs. (\ref{gmL}) and (\ref{last3}),%
\begin{equation}
\left[ p_{\alpha },p_{\beta }\right] =-\frac{i\hbar }{r^{2}}\left( L_{\alpha
\beta }+S_{\alpha \beta }\right) =-\frac{i\hbar }{r^{2}}J_{\alpha \beta }.
\label{ppj}
\end{equation}

Similarly, we can prove,%
\begin{equation}
\left[ p_{\alpha },p_{N}\right] =-\frac{i\hbar }{r^{2}}J_{\alpha N}.
\label{ppn}
\end{equation}%
Combination of two equations (\ref{ppj})\ and (\ref{ppn}) above gives the
result (\ref{ppJ}). \textit{Q.E.D.}

\section{A proof of $\left[ p_{k},J_{ij}\right] =i\hbar \left( p_{i}\protect%
\delta _{kj}-p_{j}\protect\delta _{ki}\right) $}

The proof of fundamental quantum conditions $\left[ p_{k},J_{ij}\right]
=i\hbar \left( p_{i}\delta _{kj}-p_{j}\delta _{ki}\right) $ (\ref{pJp}) is
also straightforward. To do it, we split $N$ components $p_{i}$ of the
momentum $\mathbf{p}$ into two categories of $p_{\alpha }$ $\left( \alpha
=1,2,\cdots ,N-1\right) $ and $p_{N}$. In consequence, the commutators $%
\left[ p_{k},J_{ij}\right] $ are divided into two categories as $\left[
p_{\lambda },J_{ij}\right] $ and $\left[ p_{N},J_{ij}\right] $. To process $%
\left[ p_{\lambda },J_{ij}\right] $, we compute $[p_{\lambda },J_{\alpha
\beta }]$ and $\left[ p_{\lambda },J_{iN}\right] $ one by one. The detailed
steps of calculation of commutators $[p_{\lambda },J_{\alpha \beta }]$ is
given in the following. 

We split commutators $[p_{\lambda },J_{\alpha \beta }]$ into four parts, 
\begin{equation}
\lbrack p_{\lambda },J_{\alpha \beta }]=[\Pi _{\lambda }-A_{\lambda
},L_{\alpha \beta }+S_{\alpha \beta }]=[\Pi _{\lambda },L_{\alpha \beta
}]+[\Pi _{\lambda },S_{\alpha \beta }]-[A_{\lambda },L_{\alpha \beta
}]-[A_{\lambda },S_{\alpha \beta }].  \label{pJc}
\end{equation}%
The first part of the commutators in the right-hand-side was already treated 
\cite{liu13} and the results are given by,%
\begin{equation}
\lbrack \Pi _{\lambda },L_{\alpha \beta }]=i\hbar (\Pi _{\alpha }\delta
_{\beta \lambda }-\Pi _{\beta }\delta _{\alpha \lambda }).  \label{1st}
\end{equation}%
The second part vanishes because of,%
\begin{equation}
\lbrack \Pi _{\lambda },S_{\alpha \beta }]=0.  \label{2nd}
\end{equation}%
The third part $[A_{\lambda },L_{\alpha \beta }]$ is with $L_{\alpha \beta
}\equiv x_{\alpha }\Pi _{\beta }-x_{\beta }\Pi _{\alpha },$%
\begin{subequations}
\begin{eqnarray}
\lbrack A_{\lambda },L_{\alpha \beta }] &=&[\frac{S_{\lambda \tau }n_{\tau }%
}{r+x_{N}},x_{\alpha }\Pi _{\beta }-x_{\beta }\Pi _{\alpha }] \\
&=&x_{\alpha }[\frac{S_{\lambda \tau }n_{\tau }}{r+x_{N}},\Pi _{\beta
}]-x_{\beta }[\frac{S_{\lambda \tau }n_{\tau }}{r+x_{N}},\Pi _{\alpha }] \\
&=&i\hbar (\frac{S_{\lambda \beta }n_{\alpha }}{r+x_{N}}-\frac{S_{\lambda
\alpha }n_{\beta }}{r+x_{N}}),  \label{third}
\end{eqnarray}%
where we used results (\ref{gma}). The fourth part $[A_{\lambda },S_{\alpha
\beta }]$ in (\ref{pJc}) can be easily carried out with use of the algebraic
relations (\ref{OK2}),
\end{subequations}
\begin{equation}
\lbrack A_{\lambda },S_{\alpha \beta }]=\frac{n_{\tau }}{r+x_{N}}[S_{\lambda
\tau },S_{\alpha \beta }]=-i\hbar (\delta _{\alpha \lambda }A_{\beta
}-\delta _{\beta \lambda }A_{\alpha }+\frac{n_{\alpha }S_{\lambda \beta }}{%
r+x_{N}}-\frac{n_{\beta }S_{\lambda \alpha }}{r+x_{N}}).  \label{4th}
\end{equation}

Combination of results for the third and fourth part (\ref{third}) and (\ref%
{4th}) yields a simple result,
\begin{subequations}
\begin{eqnarray}
-[A_{\lambda },L_{\alpha \beta }]-[A_{\lambda },S_{\alpha \beta }]
&=&-i\hbar (\frac{S_{\lambda \beta }n_{\alpha }}{r+x_{N}}-\frac{S_{\lambda
\alpha }n_{\beta }}{r+x_{N}})+i\hbar (\delta _{\alpha \lambda }A_{\beta
}-\delta _{\beta \lambda }A_{\alpha }+\frac{n_{\alpha }S_{\lambda \beta }}{%
r+x_{N}}-\frac{n_{\beta }S_{\lambda \alpha }}{r+x_{N}}) \\
&=&i\hbar (\delta _{\alpha \lambda }A_{\beta }-\delta _{\beta \lambda
}A_{\alpha }).
\end{eqnarray}%
Thus, the commutators $[p_{\lambda },J_{\alpha \beta }]$ give,
\end{subequations}
\begin{subequations}
\begin{eqnarray}
\lbrack p_{\lambda },J_{\alpha \beta }] &=&i\hbar (\Pi _{\alpha }\delta
_{\beta \lambda }-\Pi _{\beta }\delta _{\alpha \lambda })+i\hbar (\delta
_{\alpha \lambda }A_{\beta }-\delta _{\beta \lambda }A_{\alpha }) \\
&=&i\hbar (p_{\alpha }\delta _{\beta \lambda }-p_{\beta }\delta _{\alpha
\lambda }).  \label{last2}
\end{eqnarray}

Similarly, we can prove following two relations,
\end{subequations}
\begin{equation}
\lbrack p_{\lambda },J_{\alpha N}]=-i\hbar p_{N}\delta _{\alpha \lambda },
\label{last}
\end{equation}%
and%
\begin{equation}
\left[ p_{N},J_{ij}\right] =i\hbar \left( p_{i}\delta _{Nj}-p_{j}\delta
_{Ni}\right) .  \label{last1}
\end{equation}

In final, from results (\ref{last2})-(\ref{last1}), we see that the proof of
the fundamental quantum conditions (\ref{pJp}) is complete. \textit{Q.E.D.}

\section{Conclusions and discussions}

For a particle that is constrained on an $\left( N-1\right) $-dimensional
hypersphere, the classical bracket is $\left[ p_{i},p_{j}\right]
_{D}=-L_{ij}/r^{2}$. In contrast to the usual understanding of the
corresponding quantum condition remains $\left[ p_{i},p_{j}\right] =-i\hbar
L_{ij}/r^{2}$ in which $L_{ij}$ represents orbital angular momentum, we
argue that it must be replaced by $\left[ p_{i},p_{j}\right] =-i\hbar
J_{ij}/r^{2}$ in which $J_{ij}$ includes two parts, and one of them is the
usual orbital angular momentum $L_{ij}$ and another comes from the gauge
potential $f_{ij}$. In other words, we adapt the definition of momentum such
that $\left[ p_{i},p_{j}\right] =-i\hbar J_{ij}/r^{2}$ holds true. What is
more, we see that another set of the fundamental quantum conditions $\left[
p_{k},J_{ij}\right] =i\hbar \left( p_{i}\delta _{kj}-p_{j}\delta
_{ki}\right) $ comes true as well. The momentum is the geometric one which
came into our sight recently.

The present study gives us an insight into the noncommutativity between
different components of momentum. Our general conjecture is: where there is
such a noncommutativity, there is a generalized angular momentum which is
possibly associated with both the gauge field angular momentum and the
orbital one. The most familiar system is a charged particle in $E^{3}$ under
influence of uniform magnetic field $\mathbf{B}=B_{0}\mathbf{e}_{z}$, where
we have $[p_{x},p_{y}]=i\hbar qB_{0}$ in which $p_{x}$ and $p_{y}$ are $x$-
and $y$-component of the kinetic momentum $\mathbf{p}=-i\hbar \nabla -q%
\mathbf{A}$,  and $q$ is the charge. The magnetic field is a gauge field,
but where is the angular momentum? A usually overlooked fact is that the $z$%
-component of the field angular momentum is \cite{prl}, $%
L_{z}^{field}=qB_{0}\rho ^{2}/2$ with $\rho =\sqrt{x^{2}+y^{2}}$, we
immediately find that $[p_{x},p_{y}]=2i\hbar L_{z}^{field}/\rho ^{2}$. Thus,
the conjecture is true for the well-known system. With this understanding at
hand, we can boldly guess that the angular momentum part $f_{ij}$ in (\ref{J}%
) to be closely related to the angular momentum of the gauge field, which is
still under investigations.

\begin{acknowledgments}
This work is financially supported by National Natural Science Foundation of
China under Grant No. 11675051.
\end{acknowledgments}

\end{document}